\documentclass{article}
\usepackage{spconf,amsmath,graphicx}
\usepackage[utf8]{inputenc}
\usepackage{hyperref}

\usepackage{float}

\usepackage{color}
\usepackage{multirow}
\usepackage{glossaries}
\newacronym{asic}{ASIC}{Application-Specific Integrated Circuit}
\newacronym{awgn}{AWGN}{Additive White Gaussian Noise}

\newacronym{bm3d}{BM3D}{Block-Matching 3D}

\newacronym{cnn}{CNN}{Convolutional Neural Network}
\newacronym{crt}{CRT}{Cathode Ray Tube}
\newacronym{cpu}{CPU}{Central Processing Unit}
\newacronym{cr}{CR}{Character Recognition}

\newacronym{dvi}{DVI}{Digital Visual Interface}
\newacronym{dncnn}{DnCNN}{Denoising Convolutional Neural Network}
\newacronym{dnn}{DNN}{Deep Neural Network}
\newacronym{dp}{DP}{Display Port}
\newacronym{dl}{DL}{Deep Learning}

\newacronym{em}{EM}{Electro Magnetic}
\newacronym{emsec}{EMSEC}{Emission Security}
\newacronym{xai}{XAI}{eXplainable Articial Intelligence}

\newacronym{fcn}{FCN}{Fully Convolutional Network}
\newacronym{fc}{FC}{Fully Connected}
\newacronym{fpga}{FPGA}{Field Programmable Gate Array}
\newacronym{fpn}{FPN}{Fixed Pattern Noise}

\newacronym{gpu}{GPU}{Graphics Processing Unit}
\newacronym{gan}{GAN}{Generative Adversarial Networks}

\newacronym{hdmi}{HDMI}{High-Definition Multimedia Interface}

\newacronym{ipe}{IPE}{Information processing equipment}
\newacronym{ilsvrc}{ILSVRC}{ImageNet Large Scale Visual Recognition Competition}

\newacronym{lcd}{LCD}{Liquid Crystal Display}
\newacronym{lvds}{LVDS}{Low-Voltage Differential Signaling}

\newacronym{mse}{MSE}{Mean Square Error}
\newacronym{mae}{MAE}{Mean Absolute Error}
\newacronym{mwcnn}{MWCNN}{Multi-level Wavelet Convolutionnal Neural Network}

\newacronym{ocr}{OCR}{Optical Character Recognition}

\newacronym{psnr}{PSNR}{Peak Signal to Noise Ratio}

\newacronym{roi}{RoI}{Region of Interest}
\newacronym{rois}{RoIs}{Regions of Interest}
\newacronym{rf}{RF}{Radio Frequency}
\newacronym{rmse}{RMSE}{Root Mean Square Error}
\newacronym{rpn}{RPN}{Region Proposal Network}
\newacronym{rnn}{RNN}{Recurrent Neural Network}

\newacronym{sdr}{SDR}{Software-Defined Radio}
\newacronym{snr}{SNR}{Signal to Noise Ratio}
\newacronym{ssim}{SSIM}{Structure Similarity}

\newacronym{vga}{VGA}{Video Graphics Array}

\title{OpenDenoising: an Extensible Benchmark for\\ Building Comparative Studies of Image Denoisers}
\name{\begin{tabular}{c}Florian Lemarchand $^{\star}$, Eduardo Fernandes Montesuma $^{\star}$, Maxime Pelcat $^{\star}$, Erwan Nogues $^{\star \mathsection}$ \end{tabular}}

\address{$^{\star}$ Univ. Rennes, INSA Rennes, IETR - UMR CNRS 6164 \\
$^{\mathsection}$ DGA-MI, Bruz}
\begin{document}
\ninept
\maketitle
\begin{abstract}

Image denoising has recently taken a leap forward due to machine learning. However, image denoisers, both expert-based and learning-based, are mostly tested on well-behaved generated noises (usually Gaussian) rather than on real-life noises, making performance comparisons difficult in real-world conditions. This is especially true for learning-based denoisers which performance depends on training data. Hence, choosing which method to use for a specific denoising problem is difficult.

This paper proposes a comparative study of existing denoisers, as well as an extensible open tool that makes it possible to reproduce and extend the study. MWCNN is shown to outperform other methods when trained for a real-world image interception noise, and additionally is the second least compute hungry of the tested methods. To evaluate the robustness of conclusions, three test sets are compared. 
A Kendall's Tau correlation of only $60\%$ is obtained on methods ranking between noise types, demonstrating the need for a benchmarking tool.

\end{abstract}

\glsresetall
\section{Introduction}\label{sec:intro}

Image denoising, as a sub-domain of image restoration, is an extensively studied problem. The objective of a denoiser is to restore a cleaned image signal $\mathbf{\hat{x}}$ from an observation $\mathbf{y}$ considered to be a noisy or corrupted version of an original image $\mathbf{x}$. $\mathbf{y}$ is generated by a noise function $\mathbf{H}$ such that $\mathbf{y = H(x)}$. If the noise is additive, $\mathbf{H}$ is a sum and $\mathbf{y = x + \eta}$, where $\eta$ is a realisation of $\mathbf{H}$. There exist plenty of noise models to represent $\mathbf{H}$ among which examples of frequently used models are additive Gaussian, Poisson, salt and pepper or speckle noises. State of the art denoisers are constantly progressing in terms of noise elimination level~\cite{dabov_image_2007,zhang_beyond_2017,liu_multi-level_2018}. However, most techniques are tailored for and evaluated on a given noise distribution, exploiting its probabilistic properties to distinguish it from the signal of interest. On the specific case of Gaussian additive noise, current denoisers are approaching theoretical bounds~\cite{chatterjee_is_2010}.

Besides the largely addressed \textit{well-behaved} noise models, image restoration is also concerned by more complex noise distributions. While these distributions are application specific, they are real-world cases directly issued from identified technical needs such as image interception in difficult conditions. 

In a context where new methods are constantly appearing, it is challenging to fairly compare emerging methods to previous ones. Moreover, when a real-world noise needs to be eliminated, it is difficult to determine which of method is the best for the given noise characteristics. 
Even if most state of the art methods are evaluated on the de-facto standard databases (e.g. BSD~\cite{martin_database_2001}, 12 well-known images as Lenna, Cameraman, etc.), methods addressing specific noises and image types have to be evaluated on tailored databases. 

In this context, the contributions of this paper are:
\begin{itemize}
    \item An extensible and open-source benchmark for comparing image restoration methods.
    \item A comparative study of current denoisers on mixture and interception noise elimination, as a use case for the benchmark.
\end{itemize}

The paper is organized as follows. Section~\ref{sec:related_work} presents state of the art methods for image denoising as well as existing solutions to benchmark them.
Section~\ref{sec:proposal} describes the proposed benchmark. The comparative study, covering six restoration methods, is proposed in Section~\ref{sec:comparative_study}. Section~\ref{sec:conclu} concludes the paper.

\section{Related Work}\label{sec:related_work}

\subsection{Related Work on Benchmarks of Image Denoisers}



An active research domain in complex noise restoration is photograph restoration. This domain aims at removing a noise introduced by sensor hardware defects. Supervised datasets can be built by calibrating a sensor and hence obtaining pairs of clean and noisy samples. Darmstadt~\cite{plotz_benchmarking_2017} and PolyU~\cite{xu_real-world_2018} datasets are such datasets. Authors propose to use their datasets as a means for benchmarking denoising algorithms. This work is complementary to our proposed benchmark that can adapt to different datasets. 

Open-source projects have been created to benchmark denoising methods. The University of Toronto proposes a benchmark~\footnote{\url{www.cs.utoronto.ca/~strider/Denoise/Benchmark/}} to provide reproducibility for a method proposed in 2010~\cite{estrada_stochastic_2009}. This benchmark is tailored to the solution and not built to be extended.
Another unpublished benchmark exists that implements denoising as well as other restoration algorithms such as super-resolution~\footnote{\url{https://github.com/titsitits/open-image-restoration}}. The benchmark is limited to learning-based, Python-implemented and pre-trained methods. The latter limitation drastically reduces the use of such benchmark for complex noises. Indeed, most state of the art methods, when delivered trained, are trained on well-behaved noise. This paper proposes a benchmark extensible in several aspects. Indeed, the user can introduce his datasets, denoising methods, and metrics. 

\subsection{Related Work on Image Denoisers}
Image denoising techniques are as old as image sensors whose defects they counteract. Current denoising solutions are either expert-based denoisers, human crafted based on an expertise of artifacts or of statistical noise properties, or learning-based denoisers leveraging on latent image priors extracted from data.   

\gls{bm3d}~\cite{dabov_image_2007} is a state-of-the-art expert-based method for \gls{awgn} removal. When introduced in 2007, \gls{bm3d} clearly outperformed the previous state of the art TNRD method by a good margin of 2dB. \gls{bm3d} performs block matching to find patches with similar content in the image and uses collaborative filtering, thresholding and Wiener filtering into the transform domain to restore the image.

Authors of~\cite{vincent_stacked_2010} were the firsts to propose an encoding/decoding model with denoising criterion. Their proposal named stacked auto-encoder learns to map the noisy image to a latent space (encoding) and projects back the latent representation to the input space (decoding) to produce the denoised image. More recent auto-encoders have been proposed such as RED~\cite{mao_image_2016} that adds convolutionnal layers and uses skip connections to better keep image priors throughout the encoding/decoding pipeline.

Following these premises of denoising auto-encoders, several \gls{cnn} methods have emerged such as \gls{dncnn}~\cite{zhang_beyond_2017}. \gls{dncnn} exploits residual learning, i.e. it learns to isolate the noise $\mathbf{H}$ from the corrupted sample to later remove this noise instead of directly recovering the latent clean signal. \gls{mwcnn}~\cite{liu_multi-level_2018} is also \gls{cnn}-based. Its novelty lies in the symmetrical use of wavelet and inverse wavelet transforms into the contracting and expanding parts of a U-Net~\cite{ronneberger_u-net:_2015} architecture. The use of wavelet enables \textit{safe} subsampling with no information loss providing a better recovering of textures and sharp structures.

The most recent learning-based methods are less supervised, i.e. they require less noisy/clean image pairs to train. In Noise2Void~\cite{krull_noise2void_2019} and Noise2Noise\cite{lehtinen_noise2noise:_2018}, authors propose a tactic to train a denoising model using a single noise realisation. Authors introduce the idea of \textit{blind-spot masking} during training. They claim that the essential advantage of that strategy is to avoid to learn the identity due to the masking of the central value of the receptive field.

These methods are evaluated on well-behaved noises (typically \gls{awgn}) for the tests to be easily reproducible. Only Noise2Void is evaluated on medical images subject to complex noise. 
In the following, our open benchmark is proposed to assess denoisers' quality.


\section{Proposed Benchmark}\label{sec:proposal}

\begin{figure}
    \centering
    \includegraphics{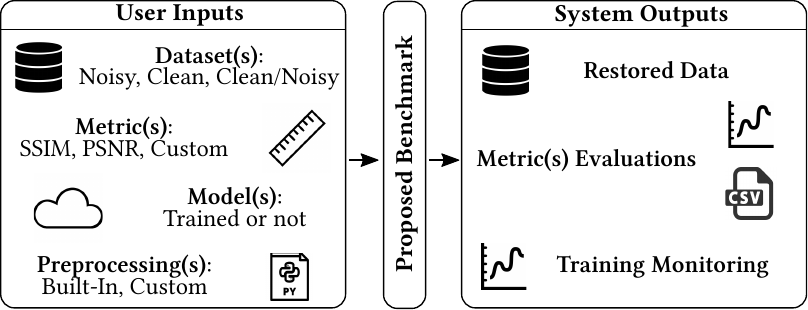}
\caption{OpenDenoising Block diagram. Users can tune datasets, metrics, denoising models and evaluation functions. OpenDenoising produces denoised samples as well as performance metrics.}
  \label{fig:schematics} 
\end{figure}

Considering the above discussed issues with existing benchmarks, we propose the OpenDenoising benchmark illustrated in Fig.~\ref{fig:schematics}. It is an open-source tool with tutorials and documentation~\footnote{\url{https://github.com/opendenoising/benchmark}} released under a CeCILL-C license.
OpenDenoising is implemented in Python and has been designed for extensions. 
Adding a new denoiser to the benchmark is a matter of minutes following a tutorial and opens for comparison with the built-in methods evoked in Section~\ref{sec:comparative_study}. For learning-based methods, the application is compatible and tested with most major frameworks (Tensorflow, Keras, Pytorch, Matlab). For learning-based training and evaluation, it is possible to use one or several datasets either supervised or not. Any scalar metric being coded in Python can be used in the benchmark. Several pre-processing functions, e.g. for data augmentation, are provided, and custom functions can be introduced. 

The user chooses whether a training is required for a method and in that case selects training parameters. Once the training is launched, monitors can be output by the benchmark to observe the learning phase. When trained models are available, evaluation is launched with custom or built-in metrics. The results are outlined using custom or built-in plots and/or stored as images or csv summaries. 

As an example of OpenDenoising versatility, it would be possible to extends the benchmark to classification methods only implementing custom evaluations metrics. 
Other potential usages of OpenDenoising include: study the extensibility of methods to new applications (see Section~\ref{sec:comparative_study}), study the strategies for re-training off-the-shelf methods (from scratch or with fine-tuning), and tune hyper-parameters. The experimental results presented in the next section exploit OpenDenoising to build a comparative study of state of the art denoisers on various types of noise.

\begin{figure}
    \centering
    \includegraphics[width=\linewidth]{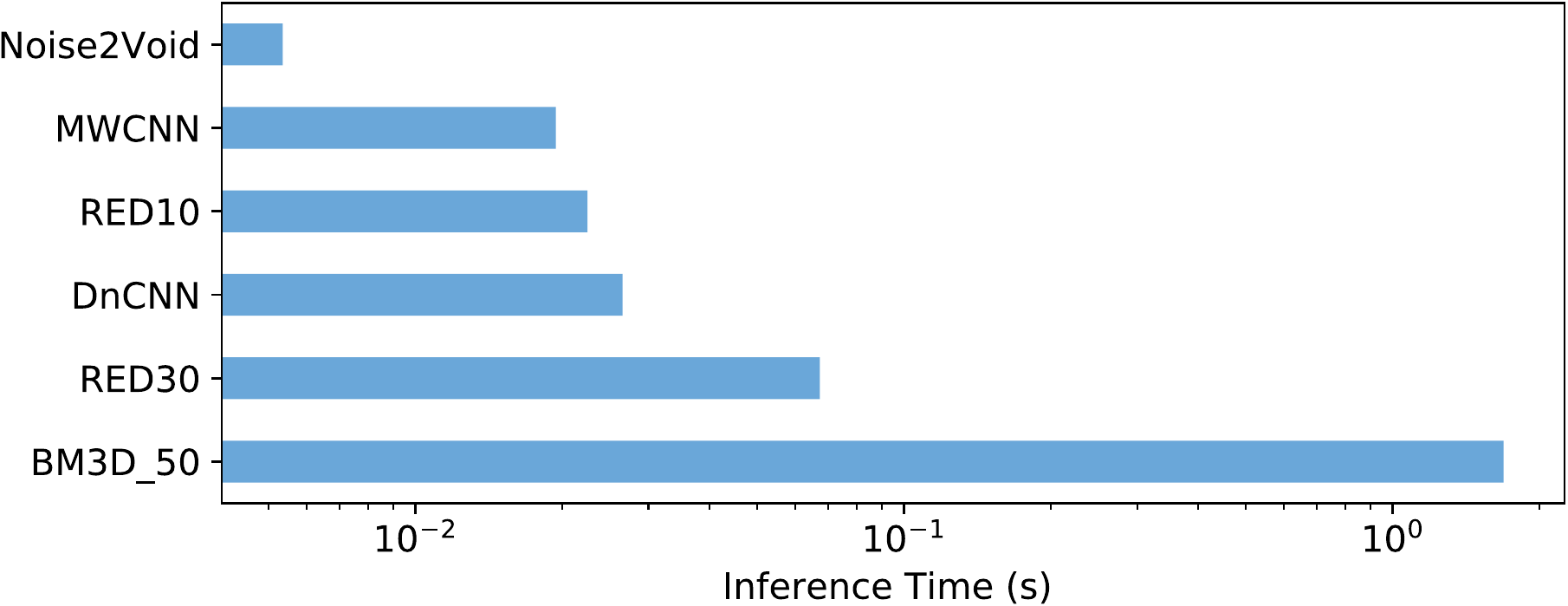}
\caption{Inference time (log scale) for different denoisers. Image resolution is $256\times256$. Noise2Void is the fastest method by almost 10 folds. \gls{mwcnn}, RED10, \gls{dncnn} and RED30 are close to each other. \gls{bm3d} is the slowest with an inference time over one second. Setup: Intel Xeon W-2125 CPU and Nvidia GTX1080 Ti GPU.}

  \label{fig:inference_barplot}     
\end{figure}

\section{A Comparative Study of Denoisers}\label{sec:comparative_study}

In this section, we apply top-ranking denoisers to images with various noises. For comparison fairness of training-based methods, no data augmentation is made and the same training datasets are used. This setup excepted, methods are trained (when applicable) using original papers parameters and training strategies. Four noise types with increasing complexity are exploited to observe the behavior of the studied denoisers. Peak Signal to Noise Ratio (PSNR) and Structural Similarity (SSIM) are used to respectively evaluate the objective and subjective quality of the denoised image.

\renewcommand{\arraystretch}{1}

\begin{table*}
    \centering
    \begin{tabular}{c l c c c c c c c}
    \multicolumn{1}{c}{} & \multicolumn{1}{c}{Dataset} & \multicolumn{1}{c}{No Denoising} & \gls{bm3d} & RED10 & RED30  & \gls{dncnn}-B & \gls{mwcnn} & Noise2Void \\
    \cline{2-9}
    \multirow{4}{*}{\rotatebox[origin=c]{90}{PSNR}} 
    & Gaussian   & $14.96$ & $23.90$  & $25.52$ & $\mathbf{25.82}$ & $25.67$ & $25.49$ & $23.41$  \\

    & Mixture   & $10.58$ & $18.29$  & $24.25$ & $\mathbf{24.58}$ & $24.50$ & $24.30$ & $19.93$  \\
 
    & Interception-Like   & $17.16$ & $22.04$  & $51.56$ & $\mathbf{52.08}$ & $51.66$ & $51.16$ & $21.70$  \\

    & Interception   & $9.46$ & $9.61$  & $22.59$ & $23.46$ & $23.04$ & $\mathbf{23.66}$ & $9.46$  \\
    \cline{2-9}
    \multirow{4}{*}{\rotatebox[origin=c]{90}{SSIM}}
     & Gaussian   & $0.24$ & $0.67$  & $0.71$ & $\mathbf{0.73}$ & $0.72$ & $0.72$ & $0.62$  \\

    & Mixture   & $0.12$ & $0.31$ & $0.67$ & $\mathbf{0.68}$ & $\mathbf{0.68}$ & $\mathbf{0.68}$ & $0.51$  \\

    & Interception-Like   & $0.11$ & $0.98$  &$\mathbf{0.99}$ &$\mathbf{0.99}$ &$\mathbf{0.99}$ &$\mathbf{0.99}$ &$0.98$  \\

    &Interception   & $0.32$ & $0.73$  & $0.94$ & $\mathbf{0.96}$ & $0.95$ & $\mathbf{0.96}$ & $0.47$  \\
    \cline{2-9}
    \end{tabular}
    \caption{Average evaluation of PSNR and SSIM metrics on test sets for, from top to bottom row: \gls{awgn} with noise level $\sigma=50$; Mixture noise made of \gls{awgn} with noise level $\sigma=50$ and salt and pepper 20\% ; Interception-Like noise being interception reference samples noised with \gls{awgn} with noise level $\sigma=50$ ; Interception noise. The test set for each noise is made of 200 samples excluded from training set.}
    \label{tab:res_bench}
\end{table*}

\subsection{Gaussian Noise}\label{subsection_gauss_noise}
\begin{figure}
    \centering
    \includegraphics[width=\linewidth]{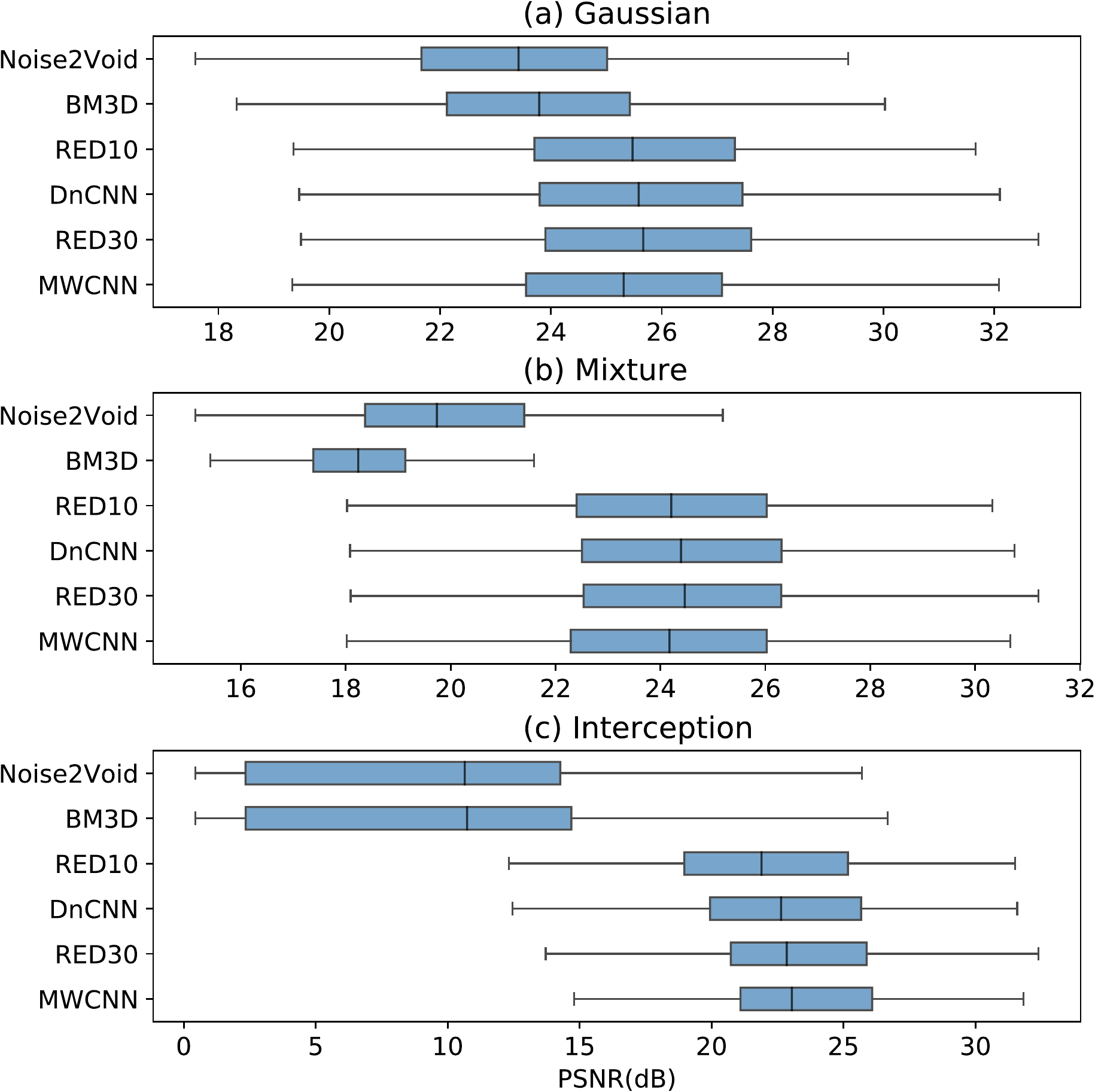}
\caption{PSNR of denoised images on (a) Gaussian noise, (b) Mixture and (c) Interception noise. Outliers are not displayed.}
  \label{fig:psnr_boxplot} 
\end{figure}

\begin{figure*}[ht]
    \centering
    \includegraphics[width=\linewidth]{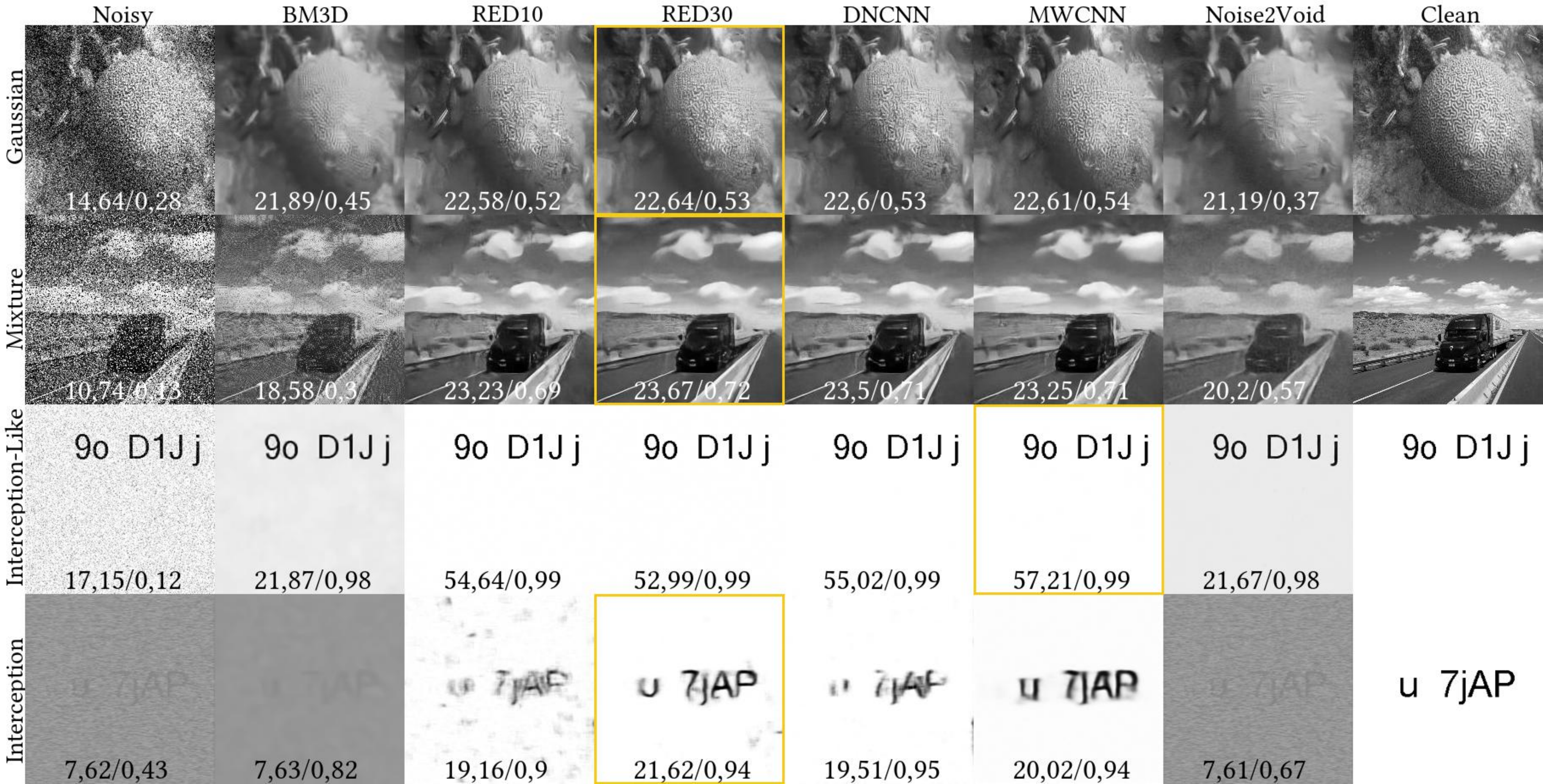}
\caption{From top to bottom, one sample per dataset is shown noisy (left), denoised with different denoisers (middle) and clean (right). PSNR/SSIM are displayed for each sample. Images with the best compromise between PSNR and SSIM metrics are yellow-boxed.}
  \label{fig:images} 
\end{figure*}

First, Gaussian noise is denoised to test the methods in their original conditions. Denoisers are evaluated on a common noisy dataset corrupted with \gls{awgn}. The underlying data is made of 10k natural images extracted from the ImageNet~\cite{deng_imagenet:_2009} evaluation set. Average PSNR and SSIM are shown in Table~\ref{tab:res_bench} and example images displayed on Fig.~\ref{fig:images}. Fig.~\ref{fig:psnr_boxplot} shows in boxplots the 10th, 25th, 75th and 90th percentiles of PSNR results as well as the median.
To focus on difficult noises, the maximum noise level commonly found in papers is picked, namely $\sigma=50$. Experimental results, shown on the first line of Table~\ref{tab:res_bench}, are coherent with the published ones, though slightly under because no data augmentation is applied. On Gaussian noise, RED30 outperforms other methods (by a limited 0.15dB $\Delta$PSNR and 1\% $\Delta$SSIM) but it is also the most costly in terms of number of parameters. 

\subsection{Mixture Noise}
Complicating the denoising task, a mixture noise is then studied. This mixture noise is constructed through the successive corruption of the samples by the previously used \gls{awgn} ($\sigma=50$) and an additional salt and pepper corruption (20\% of corrupted pixels, half 0, half maximum). The clean data is conserved and this noise mixture roughly models the behaviour of an image sensor introducing Gaussian noise because of its hardware non-uniformity and salt and pepper due to pixel defects.

Fig.~\ref{fig:psnr_boxplot} shows that learning-based methods perform consistently better than \gls{bm3d}. \gls{bm3d} is here used out of its original objective (i.e. Gaussian denoising) and thus performs poorly. Another information brought by mixture noise is that Noise2Void clearly underperforms compared to other learning-based methods. This is not surprising considering the addition of salt and pepper noise that damages the spatial coherence used as a hypothesis in the Noise2Void strategy. RED10, RED30, \gls{dncnn} and \gls{mwcnn} have close performances with a narrow victory for RED30 (0.08dB $\Delta$PSNR).

\subsection{Interception Noise}
A real-world complex noise is now studied, generated by intercepting images from \gls{em} emanations. Electronic devices produce \gls{em} emanations that not only interfere with radio devices but also compromise the data they handle. A third party performing a side-channel analysis can recover internal information from both analog~\cite{van_eck_electromagnetic_1985}, and digital circuits~\cite{kuhn_compromising_2013}. Following an eavesdropping procedure~\cite{lemarchand_electro-magnetic_2019}, it is possible to build a supervised dataset made of pairs of reference images, originally displayed on a screen, and their intercepted noisy versions. Interception strongly damages the images and denoising is necessary to interpret their content. For reproducibility, we released the dataset used for this study~\footnote{\url{https://github.com/opendenoising/interception_dataset}}. It contains more than 120k samples.

To study the noise complexity, the intercepted clean samples are also artificially corrupted using \gls{awgn} with $\sigma=50$. The resulting samples are called interception-like. As shown in Table~\ref{tab:res_bench}, most methods perform well on that denoising task. The clean content of the intercepted samples contains black characters printed on a white background. The latent clean distribution of samples is thus not an issue to denoise with learning-based methods, Noise2Void excluded. Only \gls{bm3d} and Noise2Void have problems with background restoration (see Fig.~\ref{fig:images}). This phenomena is due to the correlation of the samples content and the noising process. The background of the clean samples is fully white. When applying \gls{awgn}, half of the noise coefficients are negative and samples are clipped to an integer format. Thus, the assumption of a Gaussian distribution does not hold, leading to poor restoration results with non-supervised methods, unable to adapt. 

Table~\ref{tab:res_bench}, Fig.~\ref{fig:psnr_boxplot} and Fig.~\ref{fig:images} show the results of denoising methods applied to interception noise. Metrics drop for all methods on this complex noise. Noise2Void does not manage to denoise at all. As explained in the original paper, Noise2Void has difficulties with noise correlated between several pixels which is here the case. \gls{bm3d} is built for \gls{awgn} and is not trainable, hence the poor results in that case. Others learned methods like RED10 and \gls{dncnn} produce interesting denoising but perceptual results of Fig.~\ref{fig:images} show hardly interpretable samples, not revealed by SSIM. RED30 and \gls{mwcnn} are the best-performing methods for interception noise removal but still with some extant artifacts.

\subsection{Discussion}
Different conclusions can be drawn from the above experiments using OpenDenoising and 4 different datasets.

First, Fig.~\ref{fig:psnr_boxplot} shows that the performance ranking between methods strongly depends on noise type, training dataset and evaluation dataset. When calculating \textit{Kendall's Tau correlation coefficient}~\cite{javaid_fidelity_2010}, a value of $60\%$ is obtained between Gaussian noise ranking and Interception noise ranking, ranking based on the mean PSNR. This correlation coefficient, while being high - Kendall's Tau is in $[-1;1]$; $-1$ and $1$ respectively meaning fully discordant and fully concordant rankings -, shows the need for a benchmark such as the proposed. It automates the comparison process and the selection of a given method for a denoising problem. As an example, it would be a wrong choice to pick RED30 instead of \gls{mwcnn} for interception restoration based on the original paper evaluations (e.g. Gaussian evaluation). \gls{mwcnn} is indeed both more efficient and less computationally intensive than RED30, as shown in Fig.~\ref{fig:inference_barplot}. 

Results of Table~\ref{tab:res_bench}, Fig.~\ref{fig:psnr_boxplot} and Fig.~\ref{fig:images} show a growing gap between expert-based and learning-based method as the complexity of the denoising increase. This can be explained by the flexibility of learning-based models and the advanced information brought by a supervised training. This is evidenced by the low performance of the non-supervised Noise2Void on Mixture and Interception noises.

As stated in several studies, PSNR and SSIM do not suit well the assessment of interception restoration. Fig.~\ref{fig:images} shows that while evaluation metrics on interception noise are reasonably good (PSNR/SSIM values around $20$dB/$0.9$), the perceptual quality is poor. The explanation lies in the latent content of the samples, made of black characters on a white background. A good background restoration is sufficient to raise good evaluation metrics. This issue is evoked in~\cite{leibe_perceptual_2016} where authors propose a different evaluation metric to overcome the problem.

\section{Conclusions}\label{sec:conclu}

In this paper, the OpenDenoising tool has been proposed. OpenDenoising benchmarks image denoisers and aims at comparing methods on a common ground in terms of datasets, training parameters and evaluation metrics. Supporting several languages and learning frameworks, OpenDenoising is also extensible and open-source. The second contribution of the paper is a comparative study of image restoration in the case of a complex noise source. 

Three major conclusions arise from the comparative study. First, the difference in terms of performance between expert-based and learning-based methods rises as the complexity of the noise grows. Second, the ranking of methods is strongly impacted by the nature of the noises. Finally, \gls{mwcnn} proves to be the best method for the considered real-world interception restoration task. It slightly outperforms \gls{dncnn} and RED30 while being substantially faster.

These results show that restoring an image from a complex noise is not universally solved by a single method and that choosing a denoiser requires automated testing. Our future work includes the design of new denoising methods for intercepted images.


\newpage

\bibliographystyle{IEEEbib}
\bibliography{references}

\end{document}